\documentclass[twocolumn,tighten]{aastex62}

\usepackage{amsmath}
\usepackage{xspace}
\usepackage{multirow}
\usepackage{fancyapj}
\usepackage[modulo]{lineno}

\newcommand*\patchAmsMathEnvironmentForLineno[1]{%
  \expandafter\let\csname old#1\expandafter\endcsname\csname #1\endcsname
  \expandafter\let\csname oldend#1\expandafter\endcsname\csname end#1\endcsname
  \renewenvironment{#1}%
     {\linenomath\csname old#1\endcsname}%
     {\csname oldend#1\endcsname\endlinenomath}}% 
\newcommand*\patchBothAmsMathEnvironmentsForLineno[1]{%
  \patchAmsMathEnvironmentForLineno{#1}%
  \patchAmsMathEnvironmentForLineno{#1*}}%
\AtBeginDocument{%
\patchBothAmsMathEnvironmentsForLineno{equation}%
\patchBothAmsMathEnvironmentsForLineno{align}%
\patchBothAmsMathEnvironmentsForLineno{flalign}%
\patchBothAmsMathEnvironmentsForLineno{alignat}%
\patchBothAmsMathEnvironmentsForLineno{gather}%
\patchBothAmsMathEnvironmentsForLineno{multline}%
}

\renewenvironment{quote}%
  {\list{}{\leftmargin=0.1in\rightmargin=0.1in}\item[]}%
  {\endlist}

\newcommand{\rbr}[1]{\ensuremath{\left( #1 \right)} }
\newcommand{\sbr}[1]{\ensuremath{\left\{ #1 \right\}} }
\newcommand{\abs}[1]{\ensuremath{\left| #1 \right|} }

\newcommand{\mean}[1]{\ensuremath{<#1>} }
\newcommand{\E}[1]{\ensuremath{\times 10^{#1}} }

\newcommand{\msol}{\ensuremath{M_{\odot}}\xspace}

\newcommand{\cts}{ct\,\per{s}}
\newcommand{\per}[1]{\,#1\ensuremath{^{-1}}\xspace}
\newcommand{\persq}[1]{\,#1\ensuremath{^{-2}}\xspace}
\newcommand{\lumcgs}{erg\,\per{s}}
\newcommand{\fluxcgs}{erg\,\per{s}\,\persq{cm}}
\newcommand{\fluecgs}{erg\,\persq{cm}}

\newcommand{\rxte}{\textit{RXTE}\xspace}

\newcommand{\maxi}{\textit{MAXI}\xspace}
\newcommand{\swift}{\textit{Swift}\xspace}

\newcommand{\nicer}{\textit{NICER}\xspace}
\newcommand{\integral}{\textit{INTEGRAL}\xspace}
\newcommand{\src}{IGR~J17379\xspace}

\begin{document}

\title{On the curious pulsation properties of the accreting millisecond pulsar IGR J17379--3747}

\author{Peter Bult}
\affiliation{Astrophysics Science Division, 
  NASA's Goddard Space Flight Center, Greenbelt, MD 20771, USA}

\author{Craig B. Markwardt}
\affiliation{Astrophysics Science Division, 
  NASA's Goddard Space Flight Center, Greenbelt, MD 20771, USA}

\author{Diego Altamirano}
\affiliation{Physics \& Astronomy, University of Southampton, 
  Southampton, Hampshire SO17 1BJ, UK}

\author{Zaven Arzoumanian} 
\affiliation{Astrophysics Science Division, 
  NASA's Goddard Space Flight Center, Greenbelt, MD 20771, USA}

\author{Deepto Chakrabarty}
\affil{MIT Kavli Institute for Astrophysics and Space Research, 
  Massachusetts Institute of Technology, Cambridge, MA 02139, USA}

\author{Keith C. Gendreau} 
\affiliation{Astrophysics Science Division, 
  NASA's Goddard Space Flight Center, Greenbelt, MD 20771, USA}

\author{Sebastien Guillot} 
\affil{CNRS, IRAP, 9 avenue du Colonel Roche, BP
  44346, F-31028 Toulouse Cedex 4, France} 
\affil{Universit\'e de Toulouse, CNES, UPS-OMP, F-31028 Toulouse, France}

\author{Gaurava K. Jaisawal}
\affil{National Space Institute, Technical University of Denmark, 
  Elektrovej 327-328, DK-2800 Lyngby, Denmark}

\author{Paul. S. Ray}
\affiliation{Space Science Division, Naval Research Laboratory,
  Washington, DC 20375-5352, USA}

\author{Tod E. Strohmayer} 
\affil{Astrophysics Science Division and Joint Space-Science Institute,
  NASA's Goddard Space Flight Center, Greenbelt, MD 20771, USA}

\begin{abstract}
  We report on the \textit{Neutron Star Interior Composition Explorer}
  (\nicer) monitoring campaign of the 468 Hz accreting millisecond X-ray 
  pulsar IGR~J17379--3747. From a detailed spectral and timing analysis
  of the coherent pulsations we find that they show a
  strong energy dependence, with soft thermal emission lagging
  about 640\,$\mu$s behind the hard, Comptonized emission.
  Additionally, we observe uncommonly large pulse fractions, with
  measured amplitudes in excess of 20\% sinusoidal fractional amplitude 
  across the \nicer passband and fluctuations of up to $\sim70\%$.
  Based on a phase-resolved spectral analysis, we suggest that these
  extreme properties might be explained if the source has an unusually
  favorable viewing geometry with a large magnetic misalignment angle.
  Due to these large pulse fractions, we were able to detect
  pulsations down to quiescent luminosities ($\sim5\E{33}$\,\lumcgs).
  We discuss these low-luminosity pulsations in the context
  of transitional millisecond pulsars.
\end{abstract}


\section{Introduction}
  \label{sec:intro}
  The X-ray transient IGR J17379--3747 (hereafter \src) harbors an accreting
  neutron star in a low-mass X-binary system (LMXB). It was first
  discovered through the detection of a type I X-ray burst with IBIS/ISGRI
  aboard the International Gamma-Ray Astrophysics Laboratory
  (\integral) on February 17, 2004 \citep{Chelovekov2006}. At the
  time, the source coordinates could not be precisely determined,
  leading to shifting source designations \citep{Chelovekov2006,
  Chelovekov2010} and a separate classification in the Rossi X-ray
  Timing Explorer (\rxte) archive \citep{Markwardt2008}. The source
  was ultimately cataloged as \src \citep{Bird2007, Krivonos2007},
  with the X-ray localization determined with \swift/XRT
  \citep{Krimm2008}. The source distance is not precisely known.
  Based on its location in the direction of the Galactic center
  an assumed distance of 8.5\,kpc is typically adopted.

  The analysis of archival \integral and \rxte observations showed
  that the 2004 X-ray burst of \src occurred during a 40 day
  outburst \citep{Markwardt2008, Chelovekov2010}. The earliest
  source detection was on 2014 February 14, shortly before the
  source reached a peak X-ray luminosity of $1.1\E{36}$\,\lumcgs
  (assuming an 8.5\,kpc distance). During the outburst, \src was
  not persistently visible, rather it cycled through a series of
  reflares, each lasting about a week.

  On September 2, 2008, routine monitoring with \rxte revealed
  renewed activity from \src \citep{Markwardt2008, Shaw2008}. Slightly
  brighter, with a peak X-ray luminosity of $2.3\E{36}$\,\lumcgs,
  this second outburst did not show the reflaring pattern observed
  previously. Instead, it followed a more regular and gradual decline in
  flux, with the total outburst lasting roughly $2-3$ weeks.

  It was not until March 19, 2018 that \maxi/GSC reported the source 
  had returned to outburst \citep{Negoro2018}.
  Subsequent follow-up with the \textit{Neutron Star Interior Composition
  Explorer} (\nicer) enabled the discovery of
  468\,Hz coherent X-ray pulsations \citep{Strohmayer2018b}, 
  marking \src as an accreting millisecond X-ray pulsar (AMXP) in a 1.88
  hour binary orbit. \citet{Sanna2018b} then reinvestigated the
  archival \rxte data and were able to recover pulse detections in
  both previous outbursts.

  Following an initial decline in luminosity, the source was observed
  to re-brighten around April 9, 2018 \citep{atelEijnden2018b},
  based on which we triggered more extensive follow-up observations
  with \nicer. In this work, we report on the resulting 40 day \nicer
  monitoring campaign of \src during its 2018 outburst. In section
  \ref{sec:obs} we describe characteristics of these data.  In
  section \ref{sec:results} we present our spectral and timing
  analysis of the averaged and pulsed emission, and, finally, in
  section \ref{sec:discussion} we offer an interpretation of our
  results and a discussion of the implications for this source. 
    
\section{Observations} \label{sec:obs}
    \nicer is a non-imaging X-ray telescope mounted on the
    International Space Station (ISS; \citealt{Gendreau2016}). Its 
    X-ray Timing Instrument (XTI) is a collection of 56 co-aligned
    X-ray concentrator optics and silicon drift detector pairs. These
    detectors are sensitive in the $0.2-12$\,keV energy band
    \citep{Prigozhin2012}, providing a time resolution of
    $\sim100$\,ns (rms) and an energy resolution of $\lesssim150$\,eV.
    With 52 active detectors, \nicer has a collecting area of
    $\sim1900$\,cm$^2$ at $1.5$\,keV.

    We used \nicer to observe the X-ray transient \src starting on 
    2018 March 29 (MJD 58206.7) and continued to monitor the source 
    through 2018 May 10 (MJD 58248.3). These data are available under
    ObsID $12001401nn$, where $nn$ runs from $01$ through $27$. 
    All \nicer data were processed using the \textsc{nicerdas}
    software (version V004), which is released
    as part of \textsc{heasoft} version 6.24. Initially, we applied
    standard cleaning and filtering criteria: we selected only those
    epochs where the pointing offset was $<54\arcsec$, the dark Earth
    limb angle $>30\arcdeg$, the bright Earth limb angle $>40\arcdeg$,
    and the ISS location was outside of the South Atlantic Anomaly
    (SAA). Under these criteria, we obtained $90$\,ks of useful
    exposure. 
    
    In order to account for flaring fluctuations in the X-ray
    background, we constructed a light curve in the $12-15$\,keV energy
    band \citep[see also][]{Bult2018a}. At these energies, the
    performance of the XTI is such that essentially no astrophysical
    signal is expected.  Hence, we used this light curve as a tracer for
    increased background activity. Specifically, we binned the light
    curve using an $8$-s resolution and removed all epochs where the
    count-rate was greater than $1$\,\cts. Approximately $7$\,ks of
    exposure was filtered out with this method.  

    Finally, we used the ftool \textsc{barycorr} to apply barycentric
    corrections to the cleaned data. We used the JPL DE405 Solar System
    ephemeris \citep{Standish1998} and the radio source coordinates of
    \citet{atelEijnden2018a}. We estimated the background contribution
    from \nicer observations of the \rxte blank field regions
    \citep{Jahoda2006}. No X-ray bursts were detected.

\section{Analysis \& Results} 
\label{sec:results}

\subsection{Light Curve}
    Because \src is a faint source in the \nicer band, we limit our
    analysis to the $0.4-6$\,keV energy band, where the instrument 
    is most sensitive. The $\sim40$ day \nicer light curve for this
    energy range is shown in the top panel of Figure \ref{fig:light
    curve}.  Over the first $8$ days of our \nicer campaign, the source
    flux decayed from about $12$\,\cts to below the background level of
    about 0.5\,\cts on MJD 58213. In follow-up observations collected
    four days later, however, we could again detect the source. Over the
    following 5 days, \src showed a reflare that peaked at 4\,\cts and
    lasted until MJD 58218.6, when the source again dropped below our
    background level. Continued monitoring of \src gave positive source
    detections from MJD 58236 onwards, with a notable short term
    increase in emission on MJD 58245.8. Further monitoring was limited
    by pointing constraints.

\begin{figure*}[th]
    \centering
    \includegraphics[width=\linewidth]{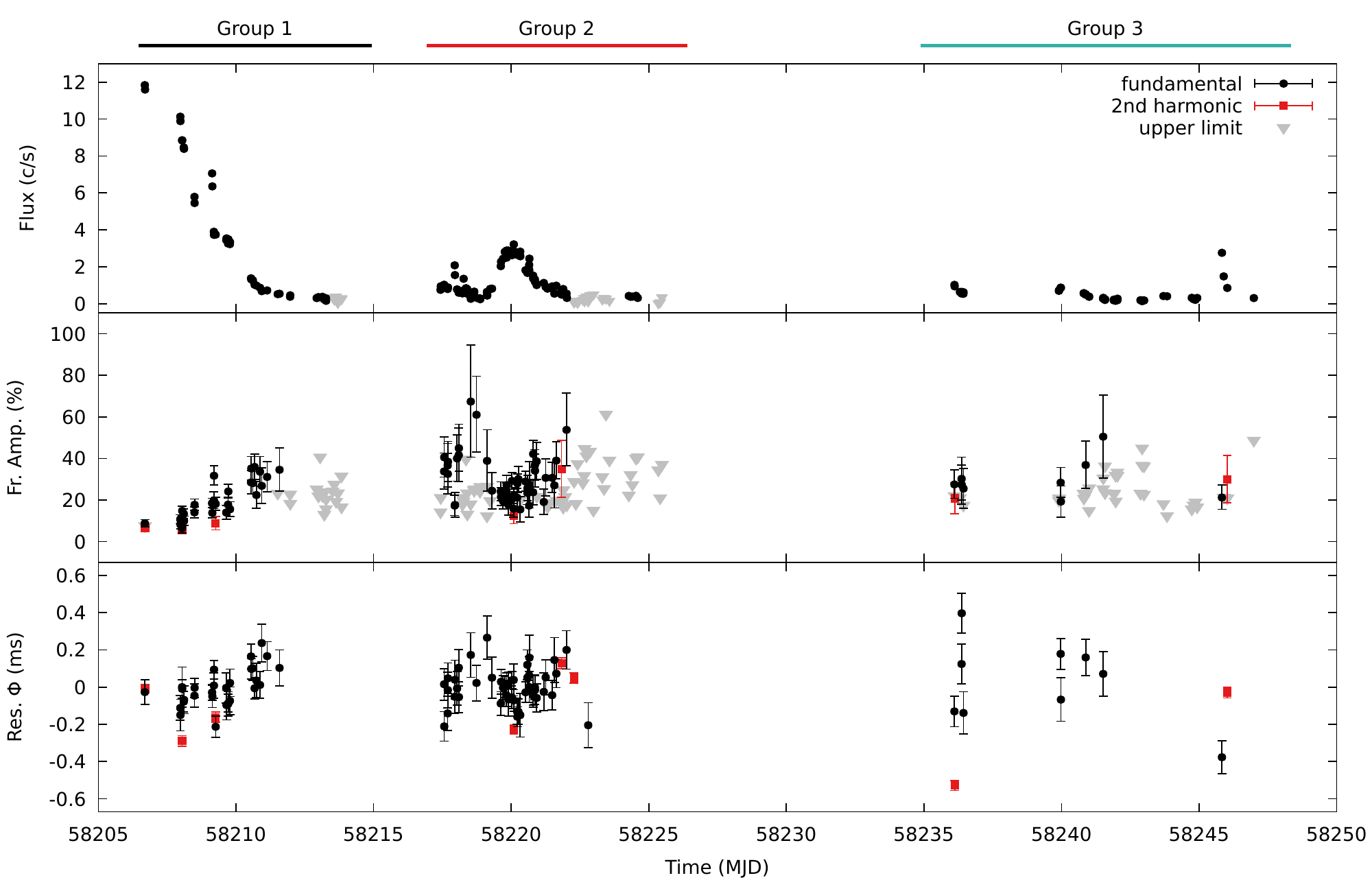}
    \caption{%
        Flux and pulse evolution of \src using $\sim300$\,s segments. 
        Top panel: background subtracted $0.4-6.0$\,keV light curve,
        with black points representing source detections and grey
        triangles giving the 95\% confidence upper limit on the source
        rate during background dominated observations. The three
        horizontal bars above the panel indicate the grouping used for
        the spectral analysis (see text and Figure \ref{fig:spectra}).
        Middle panel: fractional amplitude of the fundamental (black 
        points), the 2nd harmonic (red squares) pulsation, and the 95\% 
        confidence upper limits on the fundamental pulse.
        Bottom panel: pulse phase residuals for the best fit timing
        model (see Table \ref{tab:ephemeris}).
    }
    \label{fig:light curve}
\end{figure*}

\subsection{Swift/BAT light curve}
\label{sec:swift/bat}
\begin{figure}[t]
    \centering
    \includegraphics[width=\linewidth]{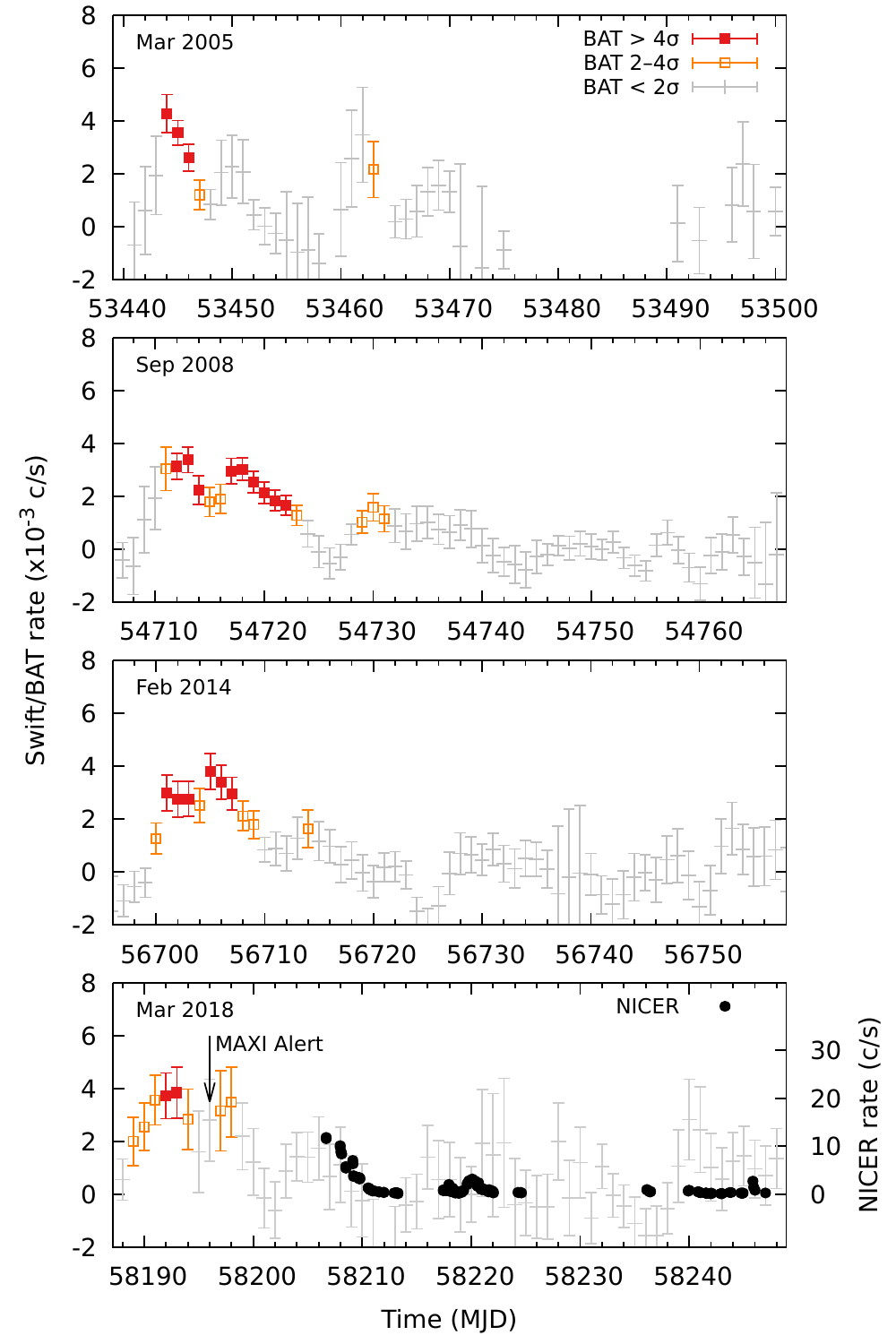}
    \caption{%
        Light curves of all historic \src outbursts as sampled with
        \swift/BAT. In all four panels data points show the 
        3-day moving average flux, with red, orange, and grey
        reflecting the source detection significance as shown in the
        legend. Additionally, bullets in the bottom panel show the
        \nicer count-rate, and the arrow indicates the date on which
        the 2018 outburst activity was first reported
        \citep{Negoro2018}.
    }
    \label{fig:swift bat}
\end{figure}

    In order to place the \nicer observations in a historic context, we 
    analysed the \swift/BAT daily monitoring light curve \citep{Krimm2013}
    to establish the past outburst behavior of \src. Hence, we applied
    a moving average to the daily light curve, using a window width of 
    3\,days and a stride of 1\,day. We then searched for all epochs
    where the averaged flux had a greater than $4\sigma$ deviation
    from zero and considered those epochs as candidate outburst
    detections.
    This approach correctly identified both the 2008 and 2018
    outbursts.  Additionally, we found there were two other epochs
    where \src appears to have been active; March 2005 and February
    2014. The light curves of all four events are shown in Figure
    \ref{fig:swift bat}. In the bottom panel of this figure, showing
    the 2018 outburst, we have further overlaid the outburst
    progression as observed with \nicer and indicated the date at
    which \maxi/GSC first reported on the 2018 outburst. The flux
    evolution of all four events is phenomenologically similar: they show a
    $1-2$ week-long main outburst cycle whose peak luminosity is clearly
    detected ($>4\sigma$), followed by a week long reflare with a peak
    flux that is only marginally detected ($2-4\sigma$). Since this
    is the same outburst pattern that was reported for the 2004
    outburst \citep{Chelovekov2006, Markwardt2008}, we argue that all
    candidates are in fact real outbursts. 

    The comparison of the \nicer and \swift/BAT light curves for the
    2018 data indicates that the initial flux decay observed with
    \nicer was likely associated with a reflare and not the first flux cycle
    of the outburst. Additionally, we see that much of the \nicer light
    curve samples source luminosities that are well below the
    detection threshold of an all-sky monitor such as \swift/BAT.
    These results indicate that \src is a much more prolific transient
    than previously believed. 

\subsection{Spectroscopy}
\label{sec:spectroscopy}
    We performed a spectroscopic analysis of our data by splitting the 
    observations into three chronological groups: the first group covers 
    the initial flux decay (up to MJD 58215), the second group
    encapsulates the reflare (up to MJD 58226), and the third group
    captures the remaining data (see also Figure \ref{fig:light curve}).

    We extracted a $0.4-6$\,keV energy spectrum for each data grouping
    and modelled those spectra using \textsc{xspec} version 12.10
    \citep{Arnaud1996}. We
    found that an absorbed Comptonized power law model provided a
    statistically adequate description of the initial flux decay
    spectrum ($\chi^2/$dof$=410/426$). Furthermore, this model did 
    significantly better than a single- or multi-temperature blackbody
    ($\chi^2/$dof$>700/426$). Due to the presence of coherent pulsations, 
    however, we may expect the source to show both thermal and
    Comptonized emission \citep[see, e.g.][]{Sanna2018b}. Indeed, we 
    found that such a two-component model provided a statistically 
    equivalent fit of the data ($\chi^2/$dof$=406/424$). We therefore
    quantified the spectrum in terms of the \textsc{xspec} model
    \begin{quote}
        \texttt{tbabs} (\texttt{bbodyrad} + \texttt{nthcomp}),
    \end{quote}
    using the abundances of \citet{Wilms2000} and cross-sections from
    \citet{Verner1996} and give the best-fit parameters in Table
    \ref{tab:spectrum}.  Additionally, we extracted a separate
    spectrum at the highest observed count-rate (ObsID $nn=01$), and
    added the \texttt{cflux} component to the above model to measure
    the $1-10$\,keV source X-ray flux (all flux measurements
    henceforth will refer to this same $1-10$\,keV range).  We found the
    highest flux to be $(4.3\pm0.1)\E{-11}$\,\fluxcgs, which translates
    to a source X-ray luminosity of $4\E{35}$\,\lumcgs, presuming a
    $8.5$\,kpc distance.
      
    The spectral group covering the reflare had a similar, but
    slightly harder, continuum shape as the first group. It could also be 
    adequately described as either a single Comptonized power law, or
    using a Comptonization plus blackbody model. We again quantified
    the spectrum in terms of the two-component model, with the
    best-fit parameters listed in Table \ref{tab:spectrum}.
    We further measured the source flux at the highest observed
    count-rate during the reflare (ObsID $nn=07$), finding a source
    flux of $(8.9\pm0.5)\E{-12}$\,\fluxcgs, which gives an approximate
    source luminosity of $7\E{34}$\,\lumcgs.

\begin{figure}[th]
    \centering
    \includegraphics[width=\linewidth]{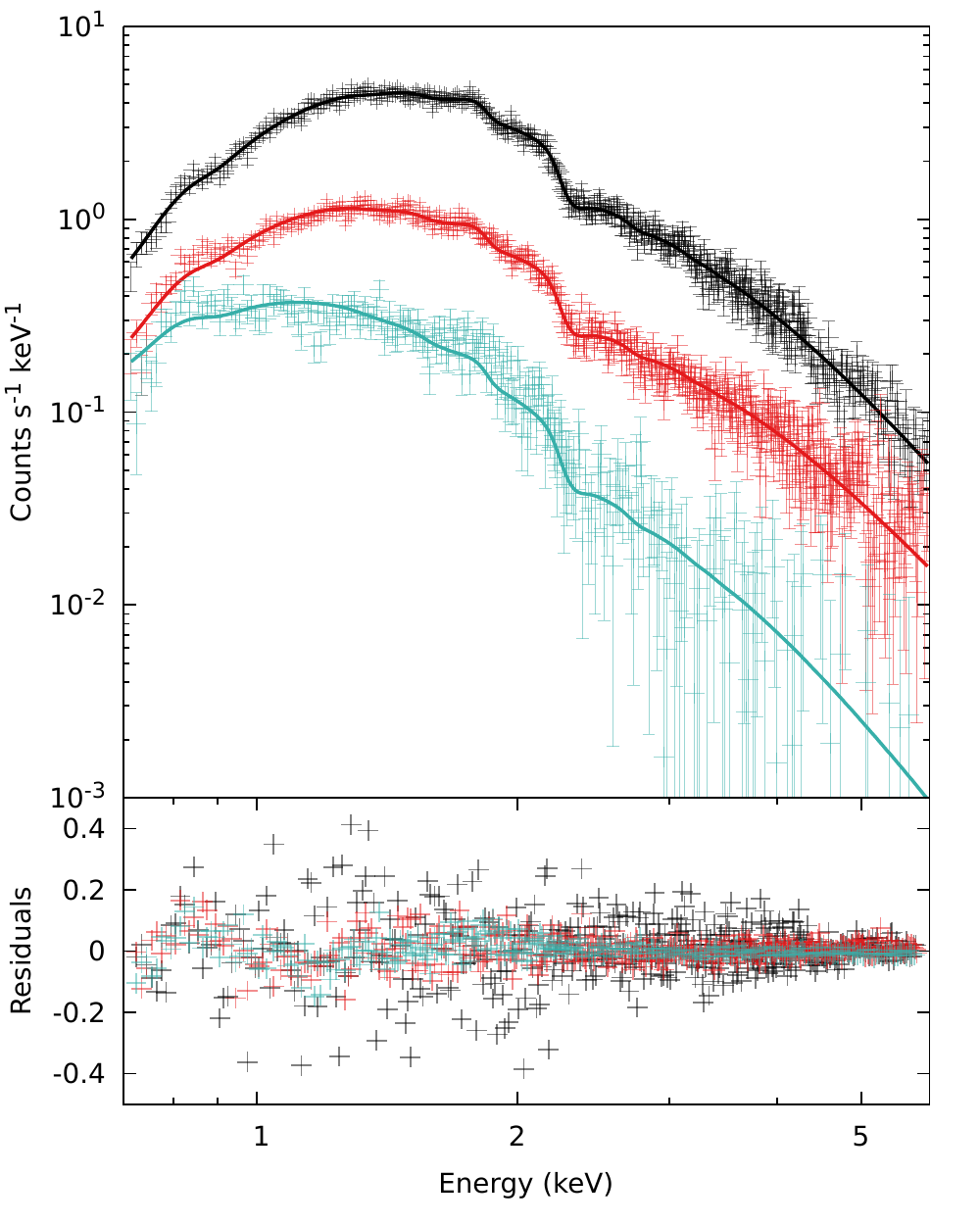}
    \caption{%
        Spectra and folded best-fit model of \src for the
        three chronological data groups respectively (1: black; 
        2: red; 3: teal; see Figure \ref{fig:light curve} and text for
        details). Top panel shows the data and best-fit model
        in units of c \per{s} \per{keV}
        and the
        bottom panel shows the best-fit residuals in the same
        units.
    }
    \label{fig:spectra}
\end{figure}

    Finally, we considered the third spectral group, which covers the
    remaining data. During this period \src mostly hovered
    just above the background level (see Figure \ref{fig:light curve}).
    The source was very soft, and could not be significantly detected
    above $\sim3$\,keV (Figure \ref{fig:spectra}). Comptonization
    models (\texttt{powerlaw} or \texttt{nthcomp}) gave a poor
    description of this spectrum and yielded unreasonably large photon
    indices ($>8$). Instead we found that the spectrum was best described
    by two blackbodies: the first at a temperature of $0.35$ keV - as
    seen in the other groups - and the second at a temperature of
    $0.12$ keV (see Table \ref{tab:spectrum} for the full model).
    We also measured the X-ray flux during this phase of prolonged low-level 
    activity, finding a flux of $(6.2\pm0.3)\E{-13}$\,\fluxcgs, with
    an associated luminosity of $5\E{33}$\,\lumcgs.

\begin{table*}
    \newcommand{\mc}[1]{\multicolumn2c{#1}}
    \newcommand{\mmc}[1]{\multicolumn3c{#1}}
    \setlength{\tabcolsep}{8pt}
    \caption{%
        Spectroscopy best fit parameters.
        \label{tab:spectrum}
    }
    \begin{center}
\begin{tabular}{l c c c c c c c c}
\tableline
~     & tbabs & \mc{bbodyrad} & \mc{bbodyrad}                  & \mmc{nthcomp}        \\
~     & $N_H$ & $kT_{bb}$ & norm & $kT_{bb}$ & norm            & $kT_{seed}$ & $\Gamma$ & norm  \\
Group & ($10^{22}$ \persq{cm}) & (keV)     & ($10^{2}$)        & (keV)
&          & (keV) & & $(10^{-3})$ \\
\tableline
1     & ~                      & - & - & $0.32\pm0.01$ & $51_{-15}^{+20}$ & $0.42\pm0.02$ & $2.7\pm0.1$ & $3.8\pm0.2$ \\
2     & $0.81_{-0.03}^{+0.06}$ & - & - & $0.38\pm0.02$ &
$6.2_{-1.3}^{+1.6}$ & $<0.13$ & $1.9\pm0.3$ & $1.5\pm0.1$ \\
3     & ~                      & $0.13\pm0.01$ & $6.8_{-2.5}^{+4.5}$ &
$0.35\pm0.02$ & $6.7_{-1.5}^{+1.9}$ & - & - & -  \\
\tableline
\end{tabular}
    \end{center}
    \tablecomments{%
        Best-fit $\chi^2$/dof = $1200 / 1090 = 1.11$.
        The \texttt{nthcomp} electron temperature was held fixed at 30
        keV and its normalizations are expressed in 
        photons \per{keV} \persq{cm} \per{s}. The \texttt{bbodyrad}
        normalization is expressed in $R_{km}^2/D_{10}^2$, with
        $R_{km}^2$ the source radius in km and $D_{10}$ the source
        distance in units of 10\,kpc. 
    }
\end{table*}

\subsection{Timing}
\label{sec:timing}
    For our timing analysis, we initially divided the $0.4-6.0$\,keV
    light curve into 32-s segments and computed a Leahy-normalized
    power spectrum for each of them. We then averaged all segments to
    a single power spectrum and renormalized the powers in terms of
    fractional rms amplitude with respect to the source flux
    \citep{Klis1995}. The resulting power spectrum showed a
    band-limited noise component at low frequencies and a distinct
    pulse spike at $468$~Hz.
    
    To further characterize the stochastic noise, we modelled the power
    spectrum using a zero-centered Lorentzian profile
    \citep{Belloni2002}. We find a goodness of fit statistic of
    $\chi^2=56$ for $67$ degrees of freedom, giving best fit
    characteristic frequency $\nu_{\rm max} = 0.32\pm0.10$\,Hz,
    and a fractional rms amplitude of $23\pm6\%$. No other
    band-limited noise or quasi-periodic variability was observed at
    either higher or lower frequencies.

    In order to characterize the coherent pulsation, we initially
    analyzed the first five ObsIDs with the aim of recovering
    the Keplerian parameters of the binary orbit. To that end, we divided 
    our $0.4-6.0$\,keV light curve into segments of continuous
    exposure (typically 600 to 1100 seconds) and applied an acceleration 
    search to each individual segment \citep{Ransom2002}. We detected the 
    pulsation in 11 of 18 segments, each giving a different centroid frequency
    and acceleration measurement. The observed frequency modulation could be 
    described by a $\nu_p=468.082$\,Hz centroid frequency pulsation
    that is shifted due to a circular orbit with period $P_b=6556$\,s,
    projected semi-major axis $a_x = 0.0706$\,lt-s, and time of
    ascending node $T_{\rm asc} = 58207.97$. This, and all reference
    times going forward, are expressed in MJD (TDB).

\subsubsection{Coherent Timing}
\label{sec:coherent timing}
    To investigate the pulse and its evolution, we performed a
    coherent timing analysis of the pulsations. For any of the
    data selections considered, we first assumed a binary ephemeris and
    adjusted the photon arrival times to the binary barycenter. Then
    we folded the data on the pulse period to construct a pulse
    profile. This profile was modeled with a constant for the
    non-pulsed contribution, and $k$ sinusoids, each with fixed
    frequency $k\nu_p$. Hence, $k=1$ described the fundamental
    pulsation, $k=2$ the second harmonic, and so forth.  An harmonic
    was considered to be significantly detected if its measured
    amplitude exceeded the $99\%$-confidence amplitude threshold of the
    noise distribution. If so, the pulse amplitude is expressed in
    terms of its sinusoidal fractional amplitude. That is,
    \begin{linenomath*}
    \begin{equation}
        r = \frac{A_k}{N_\gamma - B},
    \end{equation}
    \end{linenomath*}
    where $A_k$ is the measured amplitude of the $k$-th harmonic,
    $N_\gamma$ the number of photons in the dataset, and $B$ the
    number of photons contributed by the background emission. 
    If an harmonic amplitude was not significantly detected, we
    calculated an upper limit as the minimum signal amplitude that
    would have produced a measurement in excess of the noise threshold
    $95\%$ of the time. 

    Next, we further refined our model parameters by applying a phase
    coherent analysis to the binary period. For each segment we optimized the
    time of ascending node through a grid search method: we
    constructed a grid with varying $T_{\rm asc}$ around its
    preliminary value, propagated to be near to the observational time
    window. This grid spanned one orbital period using 1000
    steps. For each trial on the grid, we then measured the pulse
    amplitude, and picked the trial with the largest amplitude as the
    best timing solution.
    This method gave us 16 measurements of successive ascending node
    passages, which we modeled as
    \begin{linenomath*}
    \begin{equation}
        T_{{\rm asc}, k} = T_{\rm asc, ref} + P_b N_k,
    \end{equation}
    \end{linenomath*}
    where $N_k$ gives the number of orbital cycles between the
    reference epoch and the $k$-th measurement. The resulting best-fit
    parameters were\footnote{%
        The semi-major axis estimate obtained in the
        previous subsection is covariant with the binary 
        period. Hence, the improved $P_b$ measurement,
        implies an improved $a_x$.
    } 
    $T_{\rm asc} = 58209.200270$, $P_b = 
    6772.2681$\,s, with $a_x = 0.0729$\,lt-s. 
    This solution proved to be sufficiently
    accurate to allow for a coherent analysis of the pulsations.

    Finally we considered all ObsIDs. For each segment of continuous
    exposure we constructed a pulse profile and measured the pulse
    time of arrival. We fit these arrival times with a constant
    frequency and a circular orbital model using \textsc{tempo2}
    \citep{Hobbs2006} and refold the data using the improved
    ephemerides. We iterated through this procedure until the timing
    solution had converged. The best-fit parameters are listed in Table
    \ref{tab:ephemeris} and the resulting pulse amplitudes and phase
    residuals are shown in Figure \ref{fig:light curve}.  We note that our
    timing solution is statistically consistent with the long term
    timing solution reported by \citet{Sanna2018b}.

\begin{table}[t]
    \newcommand{\mc}[1]{\multicolumn2c{#1}}
    \caption{%
        Timing solution for the 2018 outburst of \src. 
        \label{tab:ephemeris}
    }
    \begin{center}
    \begin{tabular}{l l l}
        \tableline
        Parameter & {Value} & {Uncertainty} \\
        \tableline
        $\nu$ (Hz)            & 468.083266605  & 7  \E{-9} \\
        $\dot\nu$ (Hz\per{s}) &$<(-1.2 \pm 1.7)\E{-14} $ ~ \\
        $a_x \sin i$ (lt-ms)  & 76.979       & 1.4\E{-2} \\
        %$P_{b}$ (s)           & 0.07830832     & 2  \E{-8} \\
        $P_{b}$ (s)           & 6765.8388      & 1.7\E{-3} \\
        $T_{\rm asc}$ (MJD)   & 58208.966409   & 4  \E{-6} \\
        $\epsilon$            & $<5\E{-4}$     & \\ 
        \tableline
        $\chi^2$/dof          & 192 / 119      & ~ \\  
        \tableline
    \end{tabular}
    \end{center}
    \tablecomments{%
        Uncertainties give the $1\sigma$ statistical errors
        and the upper limit is quoted at the 95\% c.l. The $\epsilon$ 
        parameter gives the orbital eccentricity.
    }
\end{table}

\subsection{Energy dependence}
\begin{figure}[th]
    \centering
    \includegraphics[width=\linewidth]{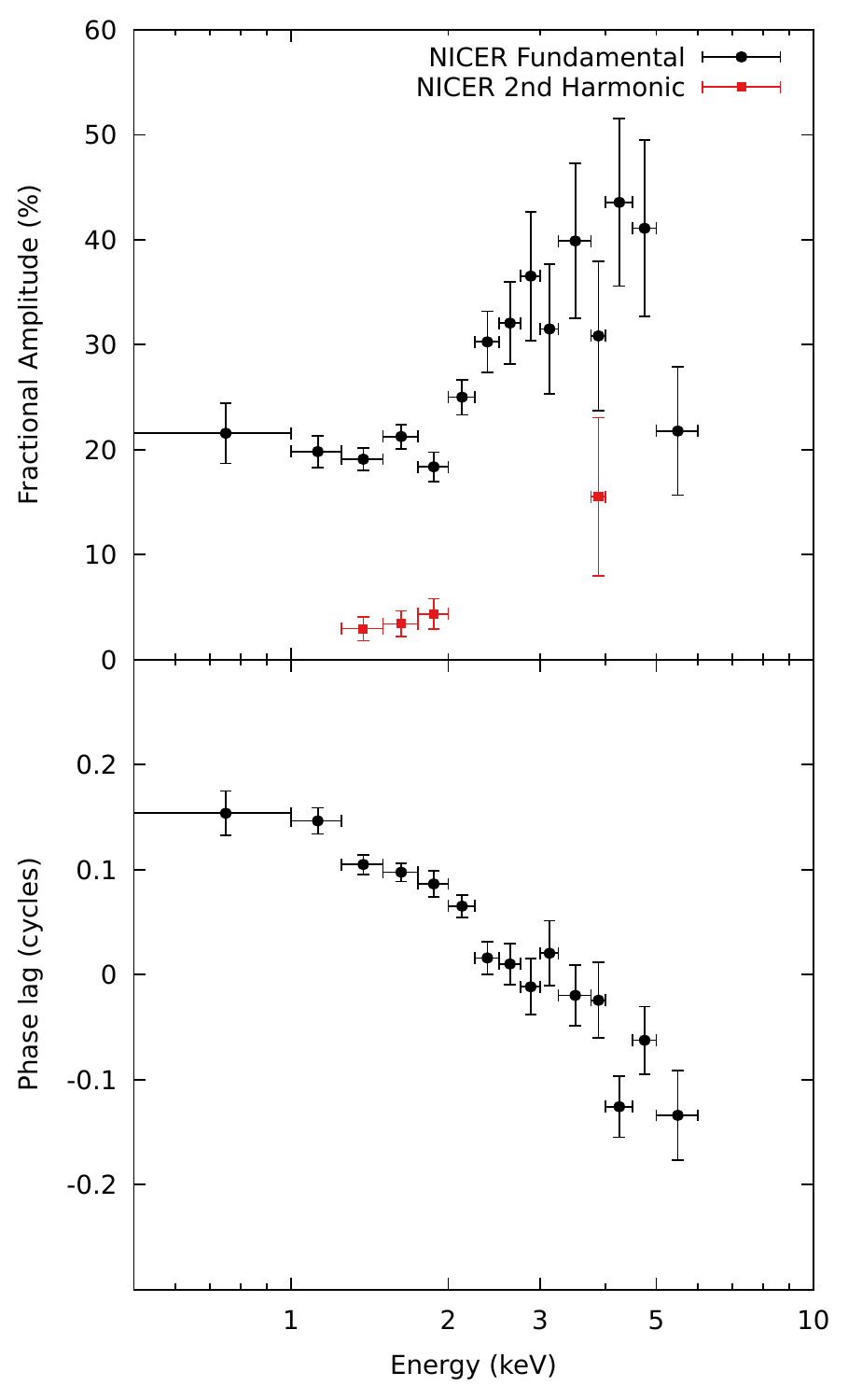}
    \caption{%
        Top: fractional pulse amplitude as a function of energy.
        Bottom: pulse phase lag with respect to the $0.4-6.0$\,keV
        timing solution as a function of energy.
    }
    \label{fig:pulse energy}
\end{figure}

    We investigated the pulse profile energy dependence by splitting
    the $0.4-6$\,keV energy range into smaller bands. For each band,
    we folded all available data on the timing solution in Table
    \ref{tab:ephemeris} and measured the pulse significance. 
    The widths of the energy bands were determined dynamically.
    Starting at the low-energy bound, we set a minimum channel width
    of 0.25\,keV. We then iteratively increased the upper bound by
    0.25\,keV until the fundamental pulse amplitude was detected at a
    significance greater than $5\sigma$, before moving on to the next
    band until the full energy range was covered. The energy-dependent
    fractional amplitudes and pulse phases resulting from this
    procedure are shown in Figure \ref{fig:pulse energy}.
    
    The fractional pulse amplitude of \src demonstrates clear energy
    dependence: below $2$\,keV the fractional amplitude is roughly
    constant at $20\%$. Above $2$\,keV the fractional amplitude
    increases rapidly to $40\%$ at 4\,keV. 
    %\nicer data does not
    %clearly establish the amplitude at higher energies, but the \nustar
    %observations suggest the factional amplitudes start decreasing.

    The pulse phase of \src also shows a clear energy dependence.
    At $0.4$\,keV the pulsations lag 0.15 cycles behind the averaged
    profile. At $6$\,keV, on the other hand, the pulsations
    \textit{lead} the averaged profile by about 0.15 cycles. Hence,
    over the \nicer passband, we observed a soft lag of about
    $640\,\mu$s, or, equivalently, of about $110\arcdeg$.
    
\subsection{Phase-resolved spectroscopy}
\label{sec:phase resolved}
  We investigated the energy dependence of the pulsations in more detail
  by performing a pulse-phase-resolved spectral analysis. We divided the pulse
  period into $8$ phase bins of equal width and extracted an energy spectrum
  for each of these bins. For this procedure, we used the same data
  grouping as described in section \ref{sec:spectroscopy}.

  Each set of phase-resolved spectra was fit using the model described
  in section \ref{sec:spectroscopy}, keeping the absorption column fixed
  at the value reported in Table \ref{tab:spectrum} and letting the
  remaining parameters vary. In our initial analysis, we found that the
  phase-dependent variations in the blackbody parameters (temperature
  and normalization) were covariant with the photon index, which
  prevented us from accurately determining either in a completely free
  fit.
  Hence, in an attempt to improve the sensitivity of our parameter
  estimation, we chose to parameterize the blackbody normalization and
  temperature using sinusoidal functions.  Specifically, we replaced the
  previously used \texttt{bbodyrad} component with a modulated
  counterpart that was defined as
  \begin{quote}
      \texttt{modbb} = \texttt{bbodyrad(kT $\times$ mod(i,r\_kt,phi\_kt))} \\
      \phantom{\texttt{modbb}} $\times$ \texttt{mod(i,r\_norm,phi\_norm)},
  \end{quote}
  where \texttt{kt} was fixed to the phase-averaged value and
  \begin{quote}
      \texttt{mod(i,r,phi)} = 1 +
      \texttt{r$\times$cos($2\pi$(i$/8-$($1/16$)+phi))},
  \end{quote}
  with \texttt{r} the fractional amplitude of the oscillation, \texttt{i} 
  fixed to the respective phase bin of the phase-resolved spectrum, and
  the $1/16$ bin offset was added to align each spectrum with the center of
  the pulse phase bin. Finally, the parameter \texttt{phi} gives the
  phase offset relative to the broad-band averaged pulse. It is defined 
  between $\pm1/2$ bin, such that a negative phase indicates a lag with
  respect to the average pulsation, and a positive phase represents a
  lead. The photon index of the \texttt{nthcomp} component was left free
  to vary with pulse phase. Because such potential variations affect the
  component spectral shape, we further chose to use a \texttt{cflux}
  component to measure variations in the integrated flux, rather than
  the normalization for the Comptonized emission. 

  Fitting our model to the phase-resolved spectra of the first spectral
  group ($\chi^2/$dof$=1591.86/1581$), we found that the blackbody
  component showed an oscillation in normalization ($r_{\rm norm} =
  19\pm8\%$), while variations in temperature were not significant.
  The Comptonized power law, meanwhile, was similarly found to
  oscillate in normalization but not in photon index. Additionally, the
  flux contribution of the power law showed a significant second harmonic. The 
  measured oscillation in flux is shown in Figure \ref{fig:phase
  resolved}, and the detailed best-fit parameters are listed in Table
  \ref{tab:phase resolved}.

  Applying the same model to the phase-resolved energy spectra of the
  second spectral group ($\chi^2/$dof$=1537.03/1590$), we found that both 
  the temperature and the normalization of the blackbody component
  showed significant oscillations at the pulse period. The temperature 
  oscillation was modest ($r_{\rm kT} = 10\pm3\%$) and compatible with
  the upper limit found previously. The normalization, on the other
  hand, was found to be consistent with being entirely pulsed ($r_{\rm
  norm} > 62\%$). Because a pulse fraction this large suggests that the
  profile might be deviating from its sinusoidal shape, we further added
  a second harmonic to the modulating blackbody. We found marginal
  evidence for the presence of such an harmonic ($\sim2\sigma$),
  suggesting that the profile is slightly asymmetric.  The Comptonized
  power law, meanwhile, again showed harmonic content in its flux
  oscillation, and a small oscillation in photon index could also be
  measured (see Table \ref{tab:phase resolved}).

  Finally, for the third spectral group, we considered a spectral model
  that consisted of two modulated blackbody components ($\chi^2/$dof$ = 
  616.13/584$). The higher temperature blackbody (i.e., the one that is
  similar to the blackbody seen in the previous spectral groups) was
  again found to be oscillating, albeit at a smaller pulse fraction
  ($r_{\rm norm} = 34\pm12\%$). The temperature was still seen to
  oscillate, with similar parameters as found earlier. 
  The lower temperature blackbody, meanwhile, appeared to be insensitive
  to rotational phase, although we note that our upper limit on the
  fractional amplitude is not especially constraining. 

\begin{figure}[t]
    \centering
    \includegraphics[width=\linewidth]{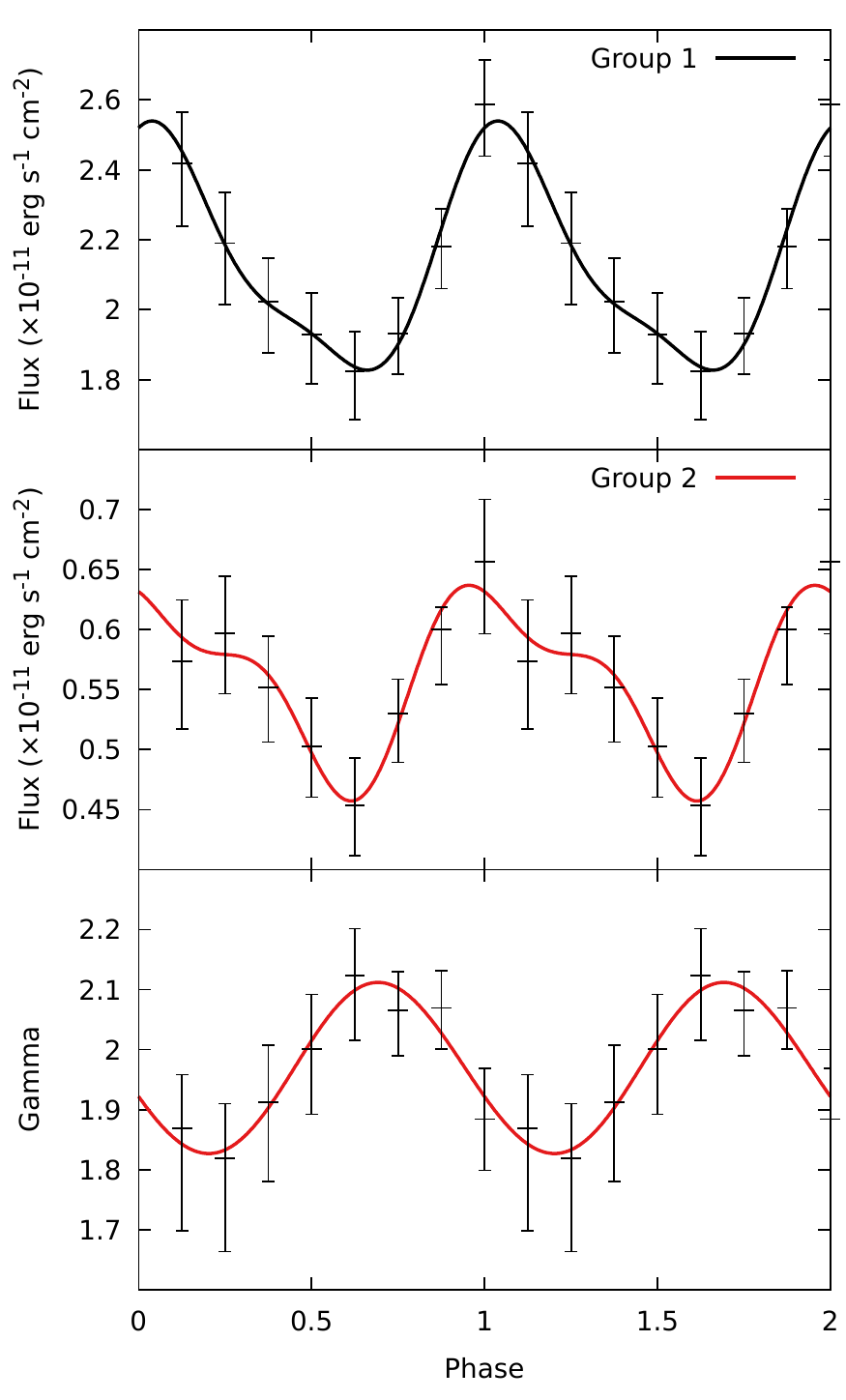}
    \caption{%
        Pulse-phase resolved variations in the Comptonized power law
        parameters. The solid lines give the best-fit harmonic
        decompositions (see Table \ref{tab:phase resolved} and text for
        details).
    }
    \label{fig:phase resolved}
\end{figure}

\begin{table}[t]
    \newcommand{\mc}[1]{\multicolumn2c{#1}}
    \newcommand{\mr}[2]{\multirow{#1}{*}{#2}}
    \caption{%
        Pulse phase resolved spectroscopy of \src. 
        \label{tab:phase resolved}
    }
    \begin{center}
    \begin{tabular}{c l l l}
        \tableline
        Group  & Parameter & {Fr. Amp.} & {Phase} \\
        \tableline
        \mr{5}{1} & modbb norm     & $(  19\pm 8 )\%$  & $     - 0.28\pm0.07$ \\
        ~         & modbb kT       & $ <13\%$          & ~ \\ %$\phm{-}0.09\pm0.18$ \\
        ~         & nthcomp Gamma  & $ <13\%$          & ~ \\
        ~         & nthcomp flux 1 & $(15.3\pm0.6)\%$  & $\phm{-}0.24\pm0.02$ \\
        ~         & nthcomp flux 2 & $( 4.3\pm0.7)\%$  & $\phm{-}0.36\pm0.03$ \\[1em]

        \mr{6}{2} & modbb norm 1   & $87^{+13}_{-25}\%$ & $\phm{-}0.01\pm0.08$ \\
        ~         & modbb norm 2   & $(22 \pm 12)\%$   & $\phm{-}0.46\pm0.12$ \\
        ~         & modbb kT       & $(10 \pm  3)\%$   & $\phm{-}0.40\pm0.04$ \\
        ~         & nthcomp Gamma  & $( 7 \pm  3)\%$   & $    {-}0.39\pm0.10$ \\
        ~         & nthcomp flux 1 & $(13.1\pm0.3)\%$  & $\phm{-}0.25\pm0.02$ \\
        ~         & nthcomp flux 2 & $( 5.9\pm0.2)\%$  & $    {-}0.41\pm0.03$ \\[1em]

        \mr{4}{3}& modbb$_{\rm low}$ norm & $ < 55\%$        & ~\\
        ~      & modbb$_{\rm low}$ kT   & $ < 11\%$          & ~\\
        ~      & modbb norm             & $34^{+10}_{-14}\%$ & $     - 0.29\pm0.06$ \\ 
        ~      & modbb kT               & $(11\pm3)\%$       & $\phm{-}0.17\pm0.09$ \\
        
        \tableline
    \end{tabular}
    \end{center}
    \tablecomments{%
        Pulsation properties for the individual spectral components
        for each of the three spectral groups (section
        \ref{sec:spectroscopy}). See Table \ref{tab:spectrum} for the
        time-averaged spectral parameters. The component
        numbering refers to the pulse harmonic, i.e., `flux 1' gives the
        fundamental pulsation in the component flux, while `flux 2'
        gives the second harmonic in the component flux.
    }
\end{table}

\section{Discussion}
\label{sec:discussion}
  We have presented the results of a coherent timing and spectral
  analysis of the AMXP \src during its 2018 outburst as observed with
  \nicer. Our monitoring campaign spanned 40 days and showed a
  three-stage progression in the light curve: a linear flux decay, a
  reflare, and a very low-flux  state during which the source luminosity
  ($\sim5\E{33}$\,\lumcgs) was on par with quiescence.
  We found that the source showed significant 468\,Hz pulsations
  throughout all three stages and that the character of those pulsations
  was highly variable. The fractional pulse amplitude was found to
  change with time from as low as $10\%$ to as high as $70\%$, with
  the bulk of our observations yielding amplitudes in the
  $20-40\%$ range. In addition to variations with time, the pulsations
  were also found to vary as a function of energy. Taken on average,
  the pulse amplitude increased with energy and exhibited a strong soft
  lag, with the $0.5$\,keV emission lagging behind the $6$\,keV emission
  by $640\,\mu$s. 

  In the following, we interpret our findings in terms of the source
  properties and accretion geometry, and explore their wider
  implications for the population of accreting millisecond pulsars. 
  We first consider the pulse properties and the results of our
  phase-resolved analysis and then move on to discuss system properties
  and the accretion process in this source and the wider population. 

\subsection{Pulse properties}
  Most AMXPs have fractional pulse amplitudes on the order of
  $1-10\%$ and show amplitude variations with time and energy 
  \citep{Patruno2012b}. While some sources have shown pulse
  fractions up to around $20\%$ \citep{Chou2008, Papitto2010}, the
  systematically large $20-40\%$ fractional amplitudes observed in 
  \src are highly unusual. 
  This suggests to us that the viewing geometry of this
  pulsar (the particular combination of its beaming pattern, 
  observer inclination\footnote{%
    The inclination angle of the neutron star spin axis.
  }, $i$, and magnetic alignment\footnote{
    The alignment angle, or magnetic colatitude, gives the angle
    between the neutron star magnetic and rotational axes.  
  }, $\theta$) is uncommonly favourable. For instance, the magnetic
  alignment angle might be large, while its offset from the inclination
  angle ($\abs{i-\theta}$) might be small. These two conditions would
  create a large amplitude aspect variation of the hot spot as the star
  rotates. 
  
  A second feature of the pulse amplitude behavior is the large range of their
  fluctuations with time (see Figure \ref{fig:light curve}). Our
  earliest observations yielded pulse fractions on the order of $10\%$,
  but the pulsations increased in amplitude and variability as the mass
  accretion rate declined, and subsequently as the source showed its
  reflare. Such fluctuations in the pulse amplitude can arise in two
  ways. 
  First, there may be a non-pulsed source of emission 
  contributing to the total flux, such as direct disk emission or
  perhaps due to lateral accretion onto the star through interchange
  instabilities \citep{Arons1976}. Both of these emission mechanisms are
  sensitive to the mass accretion rate and would cause a smaller
  non-pulsed contribution as the accretion rate drops, qualitatively
  matching our data.
  Second, fluctuations in the mass accretion rate could influence the
  size and position of the hot spot \citep{Patruno2009b, Kulkarni2013},
  which manifests as a shift in pulse amplitude and phase. Some evidence
  for this second scenario can be found in the correlation between 
  pulse amplitude and phase residuals \citep{Lamb2009} and in the
  excess timing noise found in our coherent timing analysis (section
  \ref{sec:coherent timing}). In this
  interpretation, the large shifts in pulse fractions suggest that the
  inclination is likely large, so that small changes in the hot spot
  position cause large changes in the observed pulse properties. 
  
  The energy-dependent pulse-phase lags are also unusual. While
  soft lags are ubiquitous in AMXPs \citep{Gierlinski2002, Falanga2005a,
  Gierlinski2005, Falanga2005b, Falanga2012, Sanna2017b}, the magnitude
  of these lags are much larger in \src than in other AMXPs. 
  The pulse properties of AMXPs are generally interpreted in terms of a
  two-component model, with blackbody emission originating from a thermal
  hot spot on the stellar surface, and Comptonized emission emerging from
  a shock in the accretion column above the surface
  \citep{Gierlinski2002}. Each component has a separate aspect variation
  and beaming pattern, which, together, naturally explains both the
  soft-lag and the difference in harmonic content as a function of
  energy \citep{Poutanen2003}.  A very large soft lag can then be
  explained if the Comptonized emission originates from a radially
  extended region above the hot spot \citep{Wilkinson2011}. In such a
  geometry, the effective area of the soft emission peaks when the
  hot spot is pointed toward the observer. The effective area of the hard
  emission, however, peaks when the hot spot is furthest away from the
  observer and the solid angle of the accretion column is maximized.
  Because the rotational Doppler boost is maximized as the hot spot
  rotates into view, its effect on the phase of each spectral component
  is opposite: the soft emission is shifted to an earlier phase, whereas
  the hard emission is shifted toward a later phase. Hence, the
  combination of aspect variations and Doppler boosting can plausibly
  explain a $110\arcdeg$ offset between the softest and hardest
  pulsed emission.

\subsubsection{Accretion geometry}
  A robust analysis of the accretion geometry implied by the observed
  pulse shapes requires a numerical treatment of beaming, light bending
  and other relativistic effects \citep[see, e.g.,][and references
  therein]{Salmi2018}, which is well beyond the scope of this work.
  Nonetheless, we may still obtain some first-order constraints on 
  the viewing geometry from the results of our pulse-phase-resolved
  spectral analysis.

\begin{figure}[t]
  \centering
  \includegraphics[width=\linewidth]{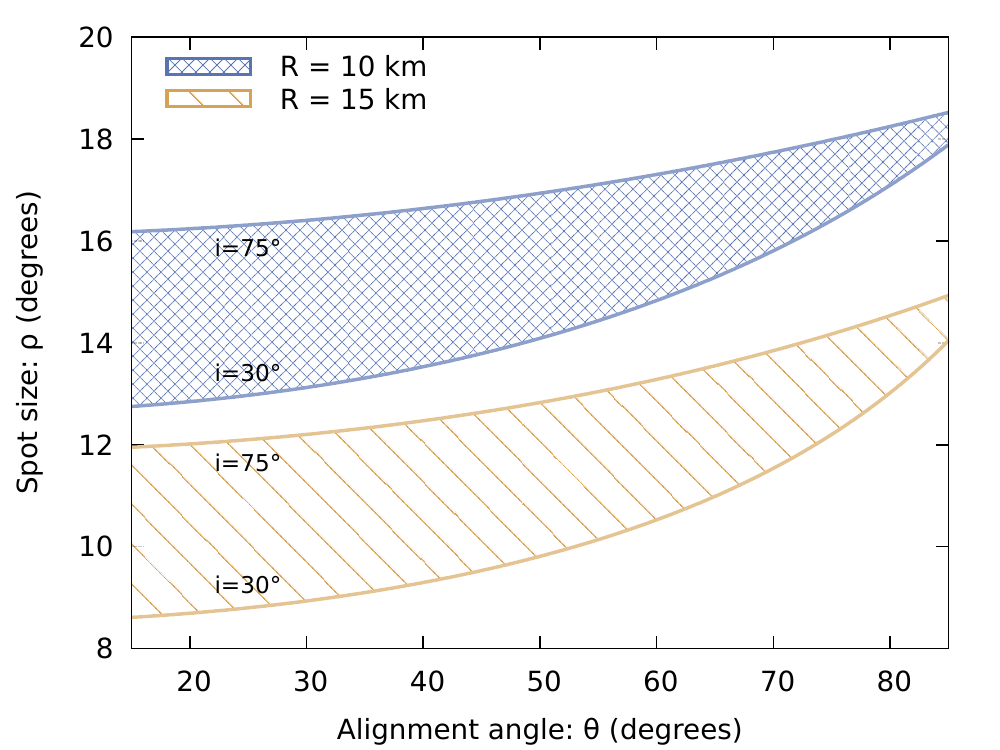}
  \caption{%
    Spot size constraints based on the observed blackbody flux
    modulation. The blue (top) band shows the range of spot sizes
    for a 10\,km neutron star radius and an inclination between
    $30\arcdeg$ and $75\arcdeg$. The brown (bottom) region show the
    same constraints for a 15\,km neutron star radius.
  }
  \label{fig:spot size}
\end{figure}

  All three of our spectral groups indicate a thermal component that
  oscillates in normalization. While the fractional amplitude of the
  oscillation varies wildly between the groups, the absolute amplitudes
  of modulation imply that the emission area shows a gradual evolution. 
  From the first to third group, we find that the apparent area of
  modulation seen at infinity is $R_{\infty} = \sbr{3.8, 2.9, 1.8}$\,km. 
  Following \citet{Poutanen2003}, we can then estimate the angular size
  of the hot spot as
  \begin{equation}
    \rho = \frac{R_\infty}{R} Q^{-1/2},
  \end{equation}
  where
  \begin{equation}
    Q = \frac{R_g}{R} + (1 - \frac{R_g}{R}) \cos i \cos \theta,
  \end{equation}
  and $R_g = 2GM/c^2$ is the gravitational Schwarzschild radius,
  with $G$ the gravitational constant, $M$ the neutron star mass, and
  $R$ its radius. In Figure \ref{fig:spot size}, we show the regions of
  allowed spot sizes as a function of $\theta$ for a choice of
  inclinations and radii, and a canonical neutron star of $1.4\msol$. 
  The inclination of \src is only weakly constrained. All we know is
  that the source does not exhibit eclipses, hence the system is not
  viewed edge on \citep{Sanna2018b}. Presuming that the neutron star rotation 
  is aligned with the orbital plane this implies $i\lesssim75\arcdeg$.
  Due to the large pulse fraction, we suggest the inclination is
  unlikely to be small, so we choose a lower bound of $30\arcdeg$. Based
  on these weak limits, we find that the spot size is likely constrained
  to $8-18\arcdeg$, independent of the alignment angle, but we note that
  larger inclinations favor larger spot sizes. 

\begin{figure}[t]
  \centering
  \includegraphics[width=\linewidth]{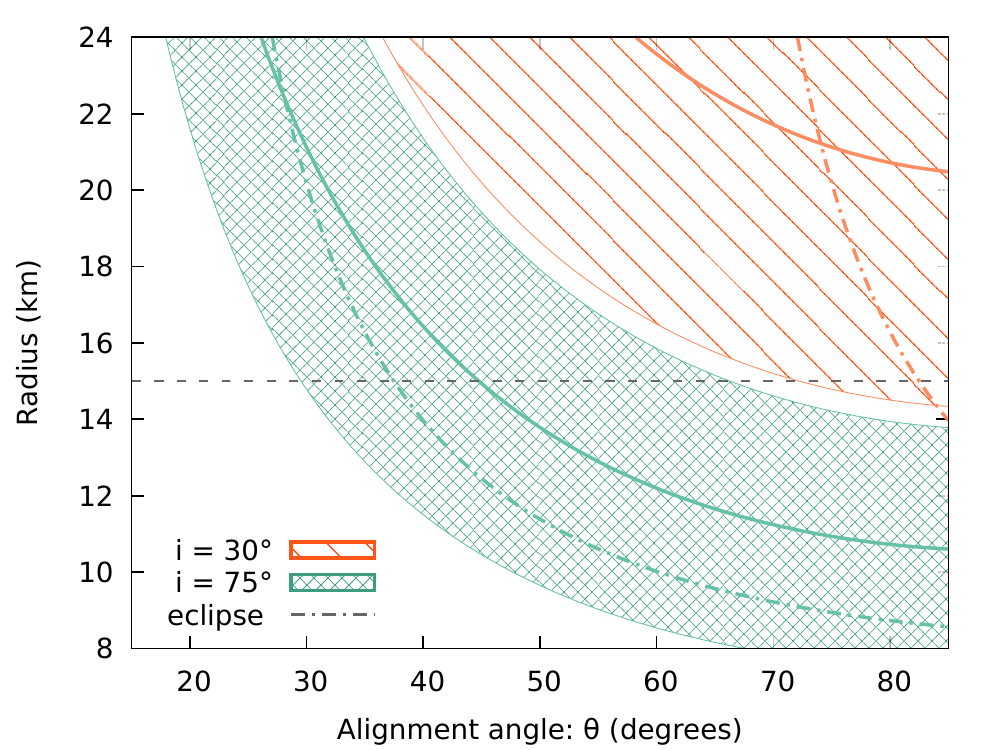}
  \caption{%
    Viewing geometry relations derived from Doppler boosting based
    on the observed $10\%$ temperature variation. The hatched areas
    mark the 90\% confidence regions. 
    The dashed-dotted line marks the hot spot eclipse boundary, such that
    for geometries to the left of the line the hot spot is always
    visible. 
    The horizontal dashed line marks a 15\,km radius to guide the eye.
  }
  \label{fig:geometry}
\end{figure}

  Two of three spectral groups also show an oscillation in blackbody
  temperature. In both cases, the oscillation fractional amplitude is on
  the order of $10\%$, which we note is consistent with the upper limit derived
  for the other spectral group. We further found that in both our
  measurements, the temperature oscillation leads the maximum in blackbody 
  area by about $150\arcdeg$. This implies that the apparent temperature
  is largest as the hot spot starts to rotate toward the observer, which
  immediately suggests an origin in a rotational Doppler boost. We
  note that the largest Doppler boost achievable is observed when the
  system is viewed edge on and the spot is located at the equator, with
  the boosting factor given by
  \begin{equation}
    \beta(R) = \frac{2\pi R \nu}{c}.
  \end{equation}
  Accounting for the alignment angle and system inclination, we can 
  relate this boosting factor to the observed change in blackbody
  temperature as \citep{Gierlinski2002}
  \begin{equation}
    \frac{kT_{\rm max}}{kT_{\rm min}} 
      \sim \frac{1 + \beta(R)\sin(i)\sin(\theta)}
                {1 - \beta(R)\sin(i)\sin(\theta)},
  \end{equation}
  so that we obtain an approximate relation between the viewing angles
  ($i$ and $\theta$) and the neutron star radius. In Figure
  \ref{fig:geometry} we plot this relation for two assumed
  inclinations and a temperature variation of $10\%$.
  While these trends are approximate and have substantial uncertainty
  regions, they still demonstrate that for realistic radii ($8-15$\,km),
  \src strongly favors large inclination ($i \gtrsim45\arcdeg$) and alignment 
  angles ($\theta \gtrsim25\arcdeg$). 
  
  In addition to a modulation of the blackbody emission parameters, we
  find that the Comptonized emission independently oscillates at the
  pulse period. Comparing the two spectral groups where this power law
  emission is present, we see that both fractional amplitude and 
  hard emission emission phase remain practically unchanged. The main
  differences are that the mean intensity goes down and that the second
  harmonic shows a phase shift, thus increasing the asymmetry of the
  hard pulse profile. Both the stability of the hard emission and the
  presence of a second harmonic support our interpretation that this
  emission originates in a radially extended region, as such a region
  would be far less sensitive to small changes in the hot spot location
  or shape. Also, different heights above the surface would experience
  slightly different Doppler boosts, which would naturally produce an
  asymmetric pulse profile \citep{Gierlinski2005}. A rather interesting
  property of our measurements is the apparent oscillation in
  photon index, which is nearly out of phase with the power law flux.
  Qualitatively, this relation might also be explained by the viewing
  geometry: the effective area of the accretion column is smallest when
  the hot spot points toward the observer, yet that is the phase at which the
  optical depth is maximized, thus the photon index should be out of
  phase. 
  
  Perhaps the most striking result of our pulse-phase-resolved analysis
  comes from the second spectral group, where we find a $>62\%$
  fractional amplitude for the blackbody normalization, and an
  apparently asymmetric profile (from the evidence for a second 
  harmonic). While we caution that this measurement has substantial
  statistical and modeling uncertainty (see Table \ref{tab:phase
  resolved} and section \ref{sec:phase resolved}), this apparent deviation of the
  pulse profile could indicate that the hot spot is being partially
  obscured by the neutron star. According to \citet{Beloborodov2002},
  such partial obscuration occurs when
  \begin{equation}
    \cos\rbr{i+\theta} < \frac{R_g}{R_g-R},
  \end{equation}
  which we have drawn as dashed-dotted lines in Figure
  \ref{fig:geometry}. We find that the condition to partial obscuration of
  the hot spot are indeed broadly consistent with the geometry
  requirements obtained from the blackbody temperature variation, and again
  prefer a system with a large magnetic misalignment angle. 

\subsection{Outburst evolution}
  The \nicer X-ray light curve showed three stages: an initial linear
  flux decay, a reflare, and a period of low activity.  Comparing this
  progression with the \swift/BAT daily monitoring light curve, we
  found that the observed linear flux decay was likely associated with a
  reflare rather than the initial outburst cycle. The initial outburst
  cycle, instead, appears to have peaked about 10 days prior (around the
  time of the \maxi/GSC alert, see Figure \ref{fig:swift bat}) and was over
  before the first \nicer observations were collected. 

\subsubsection{Reflares}
    The halting outburst progression we observed in \src is not unique to
    this source. 
    Similar reflares have been reported in other sources under various
    names, including: ``rebrightenings," ``echo-outbursts," 
    ``mini-outbursts," and ``flaring-tails". This behavior
    has been reported for dwarf novae \citep[e.g.,][]{Robertson1995,
    Kuulkers1996, Patterson2002}, Galactic black hole binary systems
    \citep[e.g.,][]{Bailyn1995, Kuulkers1996, Tomsick2004}, and
    non-pulsating accreting neutron stars \citep[e.g.,][]{Simon2010,
    Allen2015}.  Perhaps the most pertinent example of such reflaring
    behavior, however, may be found in the canonical AMXP SAX
    J1808.4--3658 \citep{Wijnands2001, Patruno2009c}.

    As in SAX J1808.4--3658, the reflares observed in \src have a
    roughly week-long duration, and peak at luminosities of
    $\sim10^{35}$\,\lumcgs. This similarity may reflect the fact that
    these two AMXPs are also very similar in terms of their orbital
    parameters and their respective stellar companions
    \citep{Chakrabarty2003,Sanna2018b}.  Important differences,
    however, exist also: the main outbursts of SAX J1808.4--3658 are
    longer than those of \src \citep{Bult2015b}.
    Additionally, SAX J1808.4--3658 is known to show a prominent aperiodic
    1-Hz modulation \citep{Klis2000,Patruno2009b} which is absent in
    \src. A very similar aperiodic 1-Hz modulation has been reported
    in the outburst of NGC 6440 X-2 \citep{Patruno2013}, a different
    AMXP whose outbursts are reminiscent of reflares in terms of
    duration and peak luminosity \citep{Altamirano2010a, Heinke2010}.
    This 1-Hz modulation has been attributed to episodic accretion onto
    the neutron star \citep{dAngelo2010, dAngelo2012}, and hence is
    likely a signature of the magnetosphere/disk interaction, rather
    than being a signature of the reflare itself. Furthermore, the 1-Hz
    modulation has also been reported to occur in the main outburst
    \citep{Bult2014}, suggesting that the underlying instability has two
    mutually exclusive branches: it occurs either around
    $10^{35}$\,\lumcgs during the reflares, or around $10^{36}$\,\lumcgs
    during the main outburst. Hence the absence of such a 1-Hz
    modulation in \src may not be meaningful. 

    It is not clear what causes some X-ray transients to show
    reflares toward the end of their outbursts. Nonetheless, as this
    phenomenon has been observed across source types, it is likely
    caused by the same ionization instability
    \citep{Osaki1974,Lasota2001} that is generally assumed to cause the
    X-ray outbursts themselves. This common interpretation essentially
    views each reflare as a separate mini-outburst: a small change in
    the disk temperature causes hydrogen to partially ionize. This,
    in turn, increases the disk opacity, which further increases its 
    temperature. As this instability grows, the mass accretion rate
    through the disk increases rapidly, and hence the source is
    observed to brighten in X-rays.
    The reason why this instability can be triggered several
    times in a row is unclear, both from an observational and
    theoretical perspective \citep{Dubus2001, Lasota2001,
    Kotko2012a}. 
    The \nicer X-ray data alone are not especially informative on
    the matter. Constraints on the origin of these
    reflares should either come from a population study or a
    multi-wavelength observing effort \citep[see
    e.g.][]{Patruno2016a}. In this light, however, we point out that
    \src appears to have a comparatively short and stable recurrence
    time and has reported counterparts in both radio
    \citep{atelEijnden2018a} and the optical/near-infrared
    \citep{Curran2011}. Hence we suggest that this source might make a
    compelling target for future multi-wavelength investigations.

\subsubsection{Low-activity/Quiescence}
    Following the reflares, we observed \src to transition into a
    prolonged low-activity state. In this state the source X-ray
    luminosity was about $5\E{33}$\,\lumcgs, placing it in the
    $\lesssim10^{34}$\,\lumcgs luminosity regime in which accreting
    neutron stars are usually considered to be in quiescence \citep{Verbunt1984}.
    In tandem with the decrease in luminosity, the source spectrum was
    observed to soften. Such late outburst spectral softening is
    commonly observed in both black hole \citep{Wu2008, Plotkin2013} and
    neutron star binaries \citep[and references therein]{Wijnands2015}.
    The soft spectrum of a quiescent neutron star can be attributed to
    either residual low-level accretion or to the gradual cooling of the
    stellar surface. It is often difficult to disentangle which of these
    two mechanisms applies, or at least, which one dominates the
    observed emission \citep[see, e.g.,][]{Fridriksson2011}. In \src,
    however, we continue to detect coherent pulsations, which is a clear
    indication that at least some channeled accretion continues even at
    the lowest luminosities. 

    Interestingly, very similar behavior has been reported for two
    of the three confirmed transitional millisecond pulsars
    \citep[tMSPs][]{Archibald2015, Papitto2015}, the class of millisecond
    pulsars that bridge the gap between traditional AMXPs and
    rotationally powered millisecond pulsars observed in the radio
    \citep{Papitto2013}. Both PSR J1023+0038 and XSS J12270--4859
    have shown coherent X-ray pulsations at luminosities of order
    $10^{33}$\,\lumcgs. In both cases, the X-ray pulsations were
    highly sinusoidal and had pulse fractions of $5-15$\%, similar to
    the X-ray pulsations seen in AMXPs. Neither of these two tMSPs,
    however, has shown the $10^{36}$\,\lumcgs X-ray outburst cycle of
    an AMXP, so the link between these two populations of pulsars has
    always been based on the similarity of their pulse profiles. 
    The third confirmed transitional millisecond pulsar, IGR
    J18245--2452, did show a high luminosity outburst
    \citep{Papitto2013}, but that outburst was highly atypical for an
    LMXB, and no X-ray pulsations have been detected at quiescent
    luminosities \citep{Linares2014}. 
    In \src, however,  we observed both the outburst cycle and the
    low-activity state, and our pulse-phase-resolved spectroscopy
    demonstrates that the character of the pulsations does not change
    substantially with luminosity. Hence, \src gives us strong evidence
    that the X-ray pulsations in the low-activity state are indeed
    accretion powered. 

\begin{figure}[t]
    \centering
    \includegraphics[width=\linewidth]{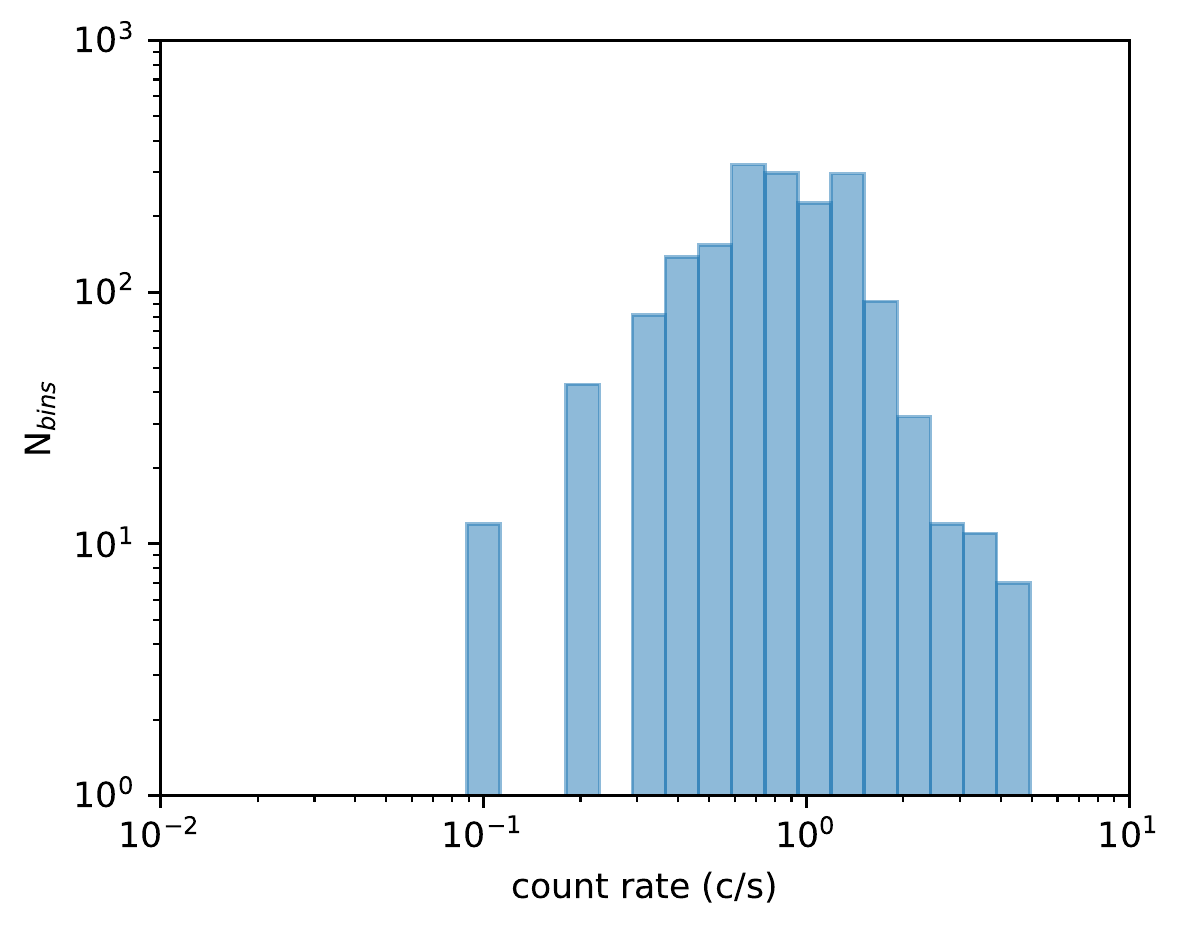}
    \caption{%
        Count rate distribution during the low-activity state of
        \src, with N$_{bins}$ giving the number of 10-s bins as a
        function of the $0.4-6$\,keV count-rate. 
    }
    \label{fig:count dist}
\end{figure}

    A second characteristic property that transitional pulsars
    show in their low activity regime, is a step-wise mode switching 
    between a ``low" and ``high" luminosity state that differs in
    observed flux by a factor of 10 \citep{Archibald2015, Bogdanov2015,
    Papitto2015}. Unlike the transitional pulsars, \src does not appear
    to be showing this characteristic mode switching.  Indeed, if we
    construct a histogram of count-rates measured in 10-s bins (Figure
    \ref{fig:count dist}), then we see no evidence of a bimodal
    population.

    A caveat to this discussion is that the distance to \src
    is not well constrained. If the source were substantially further away 
    than the presumed 8.5\,kpc, then we may be underestimating the
    luminosity by up to a factor of a few. We note, however, that
    the one X-ray burst observed from \src had a double peaked
    profile and an estimated maximum luminosity near the Eddington 
    limit \citep{Chelovekov2010}. If one assumes that this
    burst was Eddington limited \citep{Kuulkers2003}, then the
    inferred distance would place \src at $\approx8$\,kpc
    \citep{atelEijnden2018a}.  Hence, while the source distance used
    in this work is uncertain, it is probably not too far off. 
    
\subsubsection{Accretion disk configuration} 
    While the detection of X-ray pulsations at low luminosity in \src
    (and likewise in the transitional pulsars) indicates that
    magnetically channeled accretion is taking place, it is not at all clear how to
    reconcile these observations with accretion theory. In the 
    standard view of disk accretion the rotating magnetosphere of the
    neutron star imposes a centrifugal barrier \citep{Illarionov1975},
    such that accretion onto the star can proceed only while the
    disk is truncated inside the co-rotation radius
    \begin{equation}
        r_c \simeq 28~\mbox{km} 
            \rbr{\frac{\nu}{468\mbox{\,Hz}}}^{-2/3}
            \rbr{\frac{M}{1.4\,M_\odot}}^{1/3}.
    \end{equation}
    If the disk is truncated outside this co-rotation radius, then the
    propeller effect should inhibit accretion and instead drive a mass
    outflow. We can consider the radius of the magnetosphere as the
    distance at which the magnetic field is strong enough to force the
    orbiting material of a Keplerian disk into co-rotation
    \citep{Spruit1993, dAngelo2010}
    \begin{align}
        r_m &\simeq 100~\mbox{km}
            \rbr{\frac{M}{1.4\,M_\odot}}^{-1/10}
            \rbr{\frac{B}{10^8\,\mbox{Gauss}}}^{2/5}
          \nonumber \\ &\times
            \rbr{\frac{R}{10\,\mbox{km}}}^{6/5}
            \rbr{\frac{\dot M}{10^{-13}\,M_\odot\,\per{\mbox{yr}}}}^{-1/5}
          \nonumber \\ &\times
            \rbr{\frac{\nu}{468\mbox{\,Hz}}}^{-3/10},
    \end{align}
    where $B$ gives the stellar magnetic field strength and $\dot M$ the
    mass accretion rate onto the neutron star. If the observed
    $5\E{33}$\,\lumcgs luminosity is entirely due to accretion, we can
    estimate the mass accretion rate at $\dot M =
    3\E{-13}\,M_\odot\,\per{\mbox{yr}}$, which places the disk
    truncation outside the light-cylinder radius $r_{\rm lc} = 100$\,km.
    Hence, one would expect this system to be well inside the propeller
    regime, such that channeled accretion is not expected.  There are
    three scenarios proposed in the literature that could resolve this
    issue \citep[see, e.g.,][]{Archibald2015, Papitto2015, Patruno2016a}:
    \begin{itemize}
      \item % ADAF
            As the accretion rate drops, the innermost region of the
            disk may transition into a radiatively inefficient
            accretion flow, allowing the formation of an optically
            thin, geometrically thick disk \citep{Rees1982}. Such a
            disk would convert less energy into radiation, so that the
            accretion rate may be much higher than inferred from the
            X-ray flux. 
            A similar transition has been proposed to explain the
            late-outburst spectral softening in black hole binaries
            \citep{Plotkin2013}. However, it is not clear that this
            interpretation extends to neutron stars as well
            \citep{Wijnands2015}. Additionally, there is considerable 
            theoretical uncertainty surrounding the properties of such 
            a flow and how it might interact with the stellar magnetic 
            field \citep{Menou2001, DallOsso2015}, which makes it
            difficult to explore this scenario in depth.
        \item % disk outflow
            If the propeller ejects a very large fraction of the
            inflowing disk material, then it is possible that the 
            neutron star magnetosphere is experiencing a much greater
            inward pressure from the disk than what is inferred from the
            X-ray flux \citep{Lasota1999}. Plausibly, this disparity may
            be large enough to place the inner edge of the disk near
            co-rotation, thus allowing channeled accretion to proceed.
            In practice, the outflow rate would have to be roughly two
            orders of magnitude larger than the rate obtained from the
            X-ray flux, which again implies that the accretion flow must
            be radiatively inefficient, since otherwise direct emission
            from the disk would have dominated our spectrum. 
        \item % trapped disk
            Depending on the detailed microphysics governing the
            magnetosphere/disk interaction, the magnetosphere may not
            be able to drive an outflow \citep{Spruit1993}. Instead,
            the disk truncation radius could be ``trapped" near
            co-rotation \citep{Sunyaev1977,dAngelo2010}. Indeed this
            is the scenario invoked to explain the 1-Hz
            modulation in SAX J1808.4--3658 and NGC 6440 X-2
            \citep{Patruno2009c, Patruno2013}.
    \end{itemize}
    
    All three scenarios imply in one way or another that the X-ray
    flux is a poor estimate of the accretion rate onto the neutron
    star. Some independent evidence for this argument can be
    found in the parallel track phenomenon of LMXBs \citep{Klis2001}, and
    in the structured relation between the kHz QPOs and pulse amplitude
    of SAX J1808.4--3658 \citep{Bult2015b}. However, there are also
    secondary consequences to this argument. 
    
    First, if there is a large mass outflow, one might expect an
    observational signature of that outflow in the radio data. The
    detection of a flat-spectrum radio counter-part
    \citep{atelEijnden2018a} supports this scenario, although we note
    that the radio data were not contemporaneous with the low-activity
    state, and detailed modeling would be required to determine if the
    observed radio flux is consistent with the mass ejection rate required
    by our X-ray data. 

    Second, for the magnetosphere to drive an outflow, a significant
    spin-down torque would have to applied to the neutron star. The
    magnitude of this torque would have to be consistent with the long
    term spin evolution of the pulsar. The rate at which the neutron
    star spin changes due to an outflow can be estimated as
    \citep{Hartman2008}
    \begin{align}
        \dot \nu &\gtrsim -2.4\E{-13} ~ \mbox{Hz\,\per{s}} ~ n 
            \rbr{\frac{r_m}{r_c}}^{1/2}
            \rbr{\frac{I}{10^{45}\,\mbox{g\,cm}^2}}^{-1}
          \nonumber \\ &\times
            \rbr{\frac{-\dot M}{10^{-11}\,M_\odot\,\per{\mbox{yr}}}}
            \rbr{\frac{M}{1.4\,M_\odot}}^{2/3}
            \rbr{\frac{\nu_s}{468\mbox{\,Hz}}}^{-1/3}
    \end{align}
    where $n$ gives a scaling parameter capturing the detailed physics
    and $I$ is the neutron star moment of inertia.
    Roughly, we can assume $n=0$ when the disk edge is inside $r_c$,
    and $n\simeq1$ when the disk edge is in the propeller regime
    \citep{Eksi2005}. 
    Since \citet{Sanna2018b} report a lower limit on
    the long term spin frequency derivative of $\dot \nu \gtrsim
    -10^{-12}$\,Hz\,\per{s} it follows that 
    even if only $1\%$ of the accretion flow passes through 
    the barrier imposed by the propeller effect, the resulting spin-down 
    torque is still sufficiently small to be
    consistent with the observed long term spin evolution limits. We
    note, however, that the spin frequency derivative measurement for \src
    is currently limited by the relatively poor spin frequency
    measurement during the 2004 outburst \citep[see][]{Sanna2018b}. If
    a future outburst were well sampled with \nicer or a similarly
    capable timing instrument, the sensitivity to the spin frequency
    change would improve by 3 orders of magnitude. This, in turn,
    would allow for a physically interesting constraint on the
    ratio of accreted to ejected material.
    
    Finally, we note that \citet{Archibald2015} hypothesize that the mode 
    switching seen in tMSPs may be a result of the disk transitioning
    between a propeller and trapped disk state. If so, then the absence
    of mode-switching in \src may simply mean that this source is in a
    more stable trapped disk state during our observations. Similar to a
    propeller, a trapped disk must also applies a spin-down torque on the
    neutron star in order to remain stable \citep{dAngelo2010}. However,
    the loss in angular momentum for this mechanism is smaller compared
    to that predicted by mass ejection \citep{dAngelo2012}.

\subsubsection{Low-activity/Quiescent spectrum}
  The X-ray spectrum of \src during the low-activity state is
  softer than the spectra observed in the transitional
  millisecond pulsars at similar luminosities
  \citep{CotiZelati2014, Bogdanov2015, Papitto2015} and shows an
  additional blackbody component at the lowest energies. The origin
  of this blackbody is not immediately clear, although any thermal
  emission can generally be attributed to either the stellar surface
  or the accretion disk. If we interpret this low-temperature
  component as coming from the disk, then the normalization 
  gives us an implied inner disk radius of $34 (82)$\,km at $75\arcdeg$
  ($30\arcdeg$) inclination. Alternatively, 
  if the emission is originating from the neutron star, its non-pulsed
  nature suggests it has an isotropic temperature profile. It could possibly
  be generated from radiative cooling of the neutron star
  crust if the crust was heated out of equilibrium during the outburst
  \citep{Brown1998}. While the expected crust temperature
  depends on the outburst light curve, the temperature and 
  luminosity of this blackbody component matches the emission one might expect from a
  cooling neutron star crust \citep{Ootes2016}. Some of this
  uncertainty might be resolved if future outbursts could be followed
  further into quiescence to see if the source luminosity decreases over time. In order to determine
  the neutron star luminosity after cooling down from an accretion
  episode, we must first estimate the long term averaged mass accretion
  rate for this system.

\subsubsection{Long-term averaged accretion rate}
  Integrating the 2018 outburst count-rate observed with \swift/BAT,
  we find that \nicer observed roughly $10\%$ of the total outburst
  fluence.  The \nicer data, in turn, can be approximated as a
  linear decay, allowing us to measure the observed fluence as
  $8.4\E{-6}$\,\fluecgs.  Hence, we can roughly estimate the total
  fluence for this outburst at $10^{-4}$\,\fluecgs. If we assume this
  fluence to be typical for all outbursts of \src and we consider a
  recurrence time of $4$ years, we can then estimate the long-term
  averaged mass accretion rate onto the neutron star to be
  \begin{equation}
      \mean{\dot M} \simeq 6\E{-13} ~ M_\odot\,\per{\mbox{yr}}.
  \end{equation}
  This long-term averaged accretion rate is substantially lower than
  those estimated for other neutron star binaries \citep{Heinke2010}.
  If this estimate reflects the real accretion rate onto the neutron
  star, then we may expect the source to show a quiescent luminosity of
  $5\E{31}$\,\lumcgs through deep-crustal heating \citep{Brown1998}.

  Our estimate of the long-term mass accretion rate is subject to a
  number of systematic uncertainties. For one, the distance to the
  source may be larger than assumed, which would cause us to
  underestimate $\mean{\dot M}$. Additionally, \src is a faint
  source, so it is conceivable that a number of its outbursts have
  not been recorded (as evidenced by our \swift/BAT analysis, see
  section \ref{sec:swift/bat}).
  Since $\mean{\dot M}$ scales linearly with the recurrence time,
  a shorter-than-assumed recurrence would again imply that we are
  underestimating the mean accretion rate. 

  Finally, if the neutron star is less compact than assumed, e.g.,
  if its radius is 15\,km rather than the canonical 10\,km used in
  our calculations, then again the mass accretion rate is underestimated.

  However, even if we take all these uncertainties in aggregate, we
  can increase $\mean{\dot M}$ by no more than one order of magnitude. 
  Under such fine-tuning \src would still be on par with the low-end
  of the population, and have an implied quiescent luminosity of
  $\lesssim10^{32}$\,\lumcgs.

\section{Conclusions}
  Our coherent timing and spectral analysis of the AMXP \src has
  demonstrated that this source exhibits unusually large pulse fractions
  and soft phase lags. We interpreted these properties to mean that the
  source has an uncommonly favourable viewing geometry, in which the
  magnetic alignment angle is likely relatively large
  ($\gtrsim25\arcdeg$), and close to the inclination angle. The large
  pulse fraction of \src further allowed us to detect pulsations even at
  quiescent luminosities. We argued that this low luminosity state may
  be similar to the common X-ray emission state of transitional
  millisecond pulsars.
  Because of its strange pulse properties and connection to the tMSP
  population, we suggest \src is an interesting source for more detailed 
  study.

    ~\\

\nolinenumbers
\acknowledgments
This work was supported by NASA through the \nicer mission and the
Astrophysics Explorers Program, and made use of data and software 
provided by the High Energy Astrophysics Science Archive Research Center 
(HEASARC).
P.B. was supported by an NPP fellowship at NASA Goddard Space Flight Center.  
D.A. acknowledges support from the Royal Society. 

\facilities{ADS, HEASARC, NICER}
\software{heasoft (v6.24), nicerdas (v004), tempo2 \citep{Hobbs2006}}

\bibliographystyle{fancyapj}

\end{document}